# Upcoming Standards in Wireless Local Area Networks

**Sourangsu Banerji**
Department of Electronics & Communication Engineering,
RCC-Institute of Information Technology, India
Email: sourangsu.banerji@gmail.com

**ABSTRACT: In this paper, we discuss some of the upcoming standards of IEEE 802.11 i.e. Wireless Local Area Networks. The WLANs nowadays provide unlimited broadband usage to users that have been previously offered simply to wireline users within a limited range. The newest and the emerging standards fix technology issues or add functionality to the existing IEEE 802.11 standards and will be expected to overcome many of the current standing problems with IEEE 802.11.**

## 1. Introduction

The wireless broadband technologies were developed with the objective of providing services just like those supplied by, on the wireline networks. Cellular networks now provide support for top bandwidth data transfer for numerous mobile users simultaneously. In view of this, additionally, they provide mobility support for voice communication. Wireless data networks may be separated into several types depending upon their area of coverage. They are:
WLAN: Wireless Local Area network, in an area that has a cell radius up to hundred meters, mainly in office and home environments.
WMAN: Wireless Metropolitan Area Network; generally cover wider areas as huge as entire cities.
WWAN: Wireless Wide Area Network that has a cell radius about 50 km, cover areas bigger a city.

However out of every one of these standards, WLAN and recent developments in WLAN technology will be our main subject of study in this particular paper. The IEEE 802.11 is the most widely deployed WLAN technology as of today. Another renowned counterpart is the HiperLAN standard by ETSI. These two technologies are united underneath the Wireless Fidelity (Wi-fi) alliance. In literature though, IEEE802.11 and Wi-fi is used interchangeably and we will also continue with the same convention in this particular paper. A regular WLAN network is associated with an Access Point (AP) in the centre and numerous stations (STAs) are connected to this central Access Point (AP).There are just two modes in which communication normally takes place.

Within the centralized mode of communication, communication to/from a STA is actually carried across by the APs. There's also a decentralized mode in which communication between two STAs can happen directly without the requirement associated with an AP in an ad hoc fashion. WLAN networks provide coverage up to a place of 50-100 meters. Initially, Wi-fi provided an aggregate throughput of 11Mbps, but recent developments have increased the throughput to around 7 Gbps. Because of its high market penetration, several amendments on the basic IEEE 802 standard have been developed or are now under development.

In this paper, we address the technical context of its latest amendments. In section 2, we briefly discuss the history behind the development of the standard. Section 3 deals with the features of the IEEE 802 family which have already been implemented. In section 4, the upcoming standards are discussed and some open issues with the IEEE 802.11 standard is given in section 5. Lastly we conclude our paper in section 6.

## 2. Development of IEEE 802.11

The Physical layer (PHY) and medium access control (MAC) layer were mainly targeted by the IEEE 802 project. When the thinking behind wireless local area network (WLAN) was first conceived, it was just thought of another PHY of one of many available standards. The earliest candidate which was considered for it was IEEE's most prominent standard 802.3.

However later findings indicated that the radio medium behaved quite diverse from the common well behaved wire. Concerning was attenuation even over short distances, as well as collisions which could not be detected. Hence, 802.3's carrier sense multiple access with collision detection (CSMA/CD) could not be applied.

Another candidate standard considered was 802.4. At that time, its coordinated medium access i.e. the token bus concept was considered preferable over 802.3's contention-based scheme. Hence, WLAN began as 802.4L. Later in 1990 it became obvious that token handling in radio networks was rather difficult. The standardization body realized they will need a wireless communication standard that could have a very unique MAC. Finally, on March 21, 1991, the project 802.11 was approved (fig. 1).

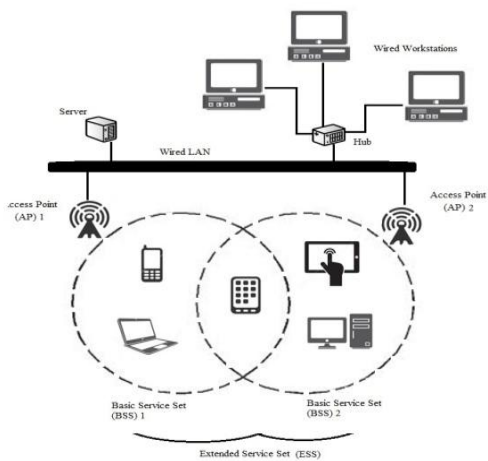

**Figure 1** WLAN Network Architecture

## 3. IEEE 802.11 family

The absolute and the most widely deployed 802.11 standard have plenty of extension and additional amendments are now under development. First introduced in 1999, the IEEE 802.11 standards were primarily developed bearing in minds our home and work environment for wireless local area connectivity. The initial standards gave a maximum data rate of 2Mbps per AP which rose up to 11 Mbps per AP with the deployment of IEEE 802.11b [2].Newer extensions like IEEE 802.11g and IEEE 802.11a provided maximum data rate of 54Mbps per AP using various methods to improve up the utmost data rates [3-5]. WLAN devices based on IEEE 802.11g currently offer data rate 100-125Mbps [4].

Similarly, a comparatively newer IEEE 802.11n provided maximum data rate around 540Mbps [25]. Current standards like IEEE 802.11ac and IEEE 802.11ad achieves data rates around 7 Gbps. Furthermore, various other standards were deployed which solved many QoS and security issues related with the older standards. Additional mechanisms were introduced to treat QoS support and security problems in IEEE 802.11e [12] and IEEE 802.11i.The IEEE 802.11n standard which we earlier discussed also introduced MAC enhancements to overcome MAC layer limitations in today's standards [28]. The IEEE 802.11s standard added mesh topology support to IEEE 802.11 [34]. The IEEE 802.11u improved internetworking with external non-802.11 networks. The IEEE 802.11w was an additional onto 802.11i covering management frame security.

The IEEE 802.11ad standard adds a "fast session transfer" feature, enabling the wireless devices to seamlessly make transition involving the legacy 2.4 GHz and 5 GHz bands and the 60 GHz frequency band [41]. The IEEE 802.11ac standard, under development is expected to supply a multi-station WLAN throughput of at the very least 7 Gbps and just one link throughput of at the very least 500 Mbps [45].

## 4. Upcoming Standards

### 4.1. IEEE 802.11ad

The IEEE 802.11ad also referred to as WiGig [fig.2] is really a relatively new standard published in December 2012. It specification adds a "fast session transfer" feature [41]. To provide for optimal performance and range criteria, the IEEE 802.11ad provides the capability to move in between the bands ensuring that computing devices are usually "best connected."

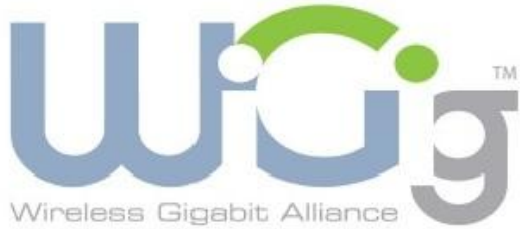

**Figure 2** WiGig Logo

Through the vast improvements in spectral reuse at 60 GHz and an efficient beam forming technology, IEEE 802.11ad enables great improvements in capacity [42]. Many users in a thick deployment can all maintain top-speed performance, without interfering with one another or having to generally share bandwidth just like the legacy frequency bands [43-44]. The likely enhancements to 802.11 beyond a brand new 60 GHz PHY include MAC modifications for directional antennas, personal basic service set, beamforming, fast session transfer between PHYs and spatial reuse.

**Table 1** Features of IEEE 802.11ad

| Parameter | Details |
|---|---|
| Operating frequency range | 60 GHz ISM band |
| Maximum data rate | 7 Gbps |
| Typical distances | 1 - 10 m |
| Antenna technology | Uses beamforming |
| Modulation formats | Various: single carrier and OFDM |

Using frequencies in the millimeter range IEEE 802.11ad microwave Wi-Fi has a range that is measured of a few meters. The aim is that it will be used for very short range (across a room) high volume data transfers such as HD video transfers [47].Table 1 summarizes the key features of this amendment.

The signal spectrum and spectral mask needs to ensure that the signal is maintained within a certain bandwidth. The spectral mask shows the mask for the spectrum (fig. 3).

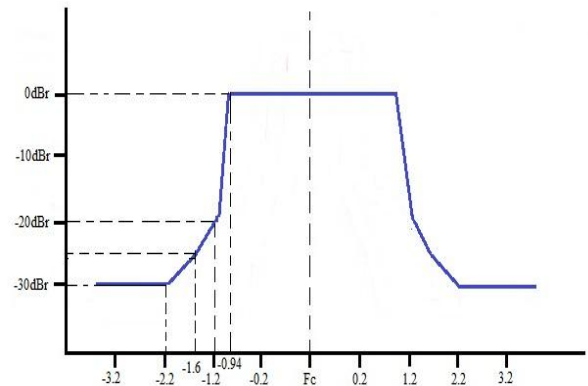

**Figure 3** IEEE 802.11ad Spectral Mask measured with 100 kHz resolution bandwidth and 30 kHz video bandwidth.

Here, dBr = dB relative to the maximum spectral density of the signal.

### 4.1.1. Physical Layer

The 802.11ad PHY supports five main signals with different modulation. These are specified as:

**i) Control PHY (CPHY)**: Providing control, this signal has high levels of error correction and detection. Accordingly it has a relatively low throughput. As it does not carry the main payload, this is not an issue. It exclusively carries control channel messages. The CPHY uses differential encoding, code spreading and BPSK modulation [46].

**ii) Single Carrier PHY:** The SCPHY employs single carrier modulation techniques: BPSK, QPSK or 16-QAM on a suppressed carrier located on the channel centre frequency. This signal has a fixed symbol rate of 1.76 Gsym/sec. A variety of error coding and error coding modes are available according to the requirements [46].

**iii) Orthogonal Frequency Division Multiplex PHY (OFDMPHY):** As with any OFDM scheme, the OFDMPHY uses multicarrier modulation to provide high modulation densities and higher data throughput levels than the single carrier modes.

**iv) Spread QPSK (SQPSK):** It involves using paired OFDM carriers onto which the data is modulated. The two carriers are maximally separated to improve the robustness of the signal in the presence of frequency selective fading.

**v) Low Power Single Carrier PHY (LPSCPHY):** This 802.11ad signal uses a single carrier as the name implies, and this is to minimize the power consumption. It is intended for small battery devices that may not be able to support the processing required for the OFDM format [46].

### 4.2. IEEE 802.11ae

The IEEE 802.11ae aims to introduce a mechanism for prioritization of management frames. A protocol to communicate management frame prioritization policy is specified in this standard.

### 4.3. IEEE 802.11ac

Among the important standards currently under development is IEEE 802.11ac. This standard is anticipated to be published by the end of 2014. It's expected to supply a multi-station WLAN throughput of around 7 Gbps and an individual link throughput of at the least 500 Mbps [45]. That is accomplished by extending the air interface concepts which are embraced by 802.11n like wider RF bandwidth (up to 160 MHz), more MIMO spatial streams (up to 8), multi-user MIMO, and high-density modulation [45-47].

IEEE 802.11ac will use OFDM like the previous 802.11n standard as the OFDM modulation scheme is particularly helpful in wideband data transmission overcoming a lot of problems faced in selective channel fading.

In order to achieve the required spectral usage of 7.5 bps/Hz, MIMO is required, and in the case of IEEE 802.11ac Wi-Fi, a form known as Multi-User MIMO or MU MIMO is also implemented. MIMO provides a number of alternate ways to utilize the number of signal paths that exist between the transmitter and the receiver to significantly improve up the data throughput. MIMO technology enables the system to set up multiple data streams on the same channel, thereby increasing the data capacity of a channel [48].

**Table 2** Features of IEEE 802.11ac

| Parameter | Details |
|---|---|
| Frequency band | 5.8 GHz ISM (unlicensed) band |
| Max data rate | 6.93 Gbps |
| Transmission bandwidth | 20, 40, & 80 MHz<br>160 & 80 + 80 MHz optional |
| Modulation formats | BPSK, QPSK, 16-QAM, 64-QAM<br>256-QAM optional |
| FEC coding | Convolutional or LPDC (optional) with coding rates of 1/2, 2/3, 3/4, or 5/6 |
| MIMO | Both single and multi-user MIMO with up to 8 spatial streams. |
| Beam-forming | Optional |

On the other hand, MU-MIMO refers to the simultaneous transfer of data frames to multiple users on the network. The IEEE 802.11ac is determined to use specialized queuing mechanisms for appropriate handling of different data which is required to be sent to more than one receiver at a time. Table 2 summarizes the key features of this amendment.

The previous versions of 802.11 standards have typically used 20 MHz channels, although 802.11n used up to 40 MHz wide channels which was optional. The 802.11ac standard aims to use channel bandwidths up to 80 MHz as standard with options of 160MHz or two 80MHz blocks. To achieve this it is necessary to adapt automatic radio tuning capabilities so that higher-bandwidth channels are only used where necessary to conserve spectrum. Table 3 summarizes the PHY Layer specifications.

Like the previous transmission standards, 802.11ac is also given a spectral mask into which the emitted signals must fall. This spectral mask details the maximum level of spurious signals and noise that are permissible.

The spectral mask differs between the various bandwidths and also according to the offset from the centre frequency. [fig. 4]

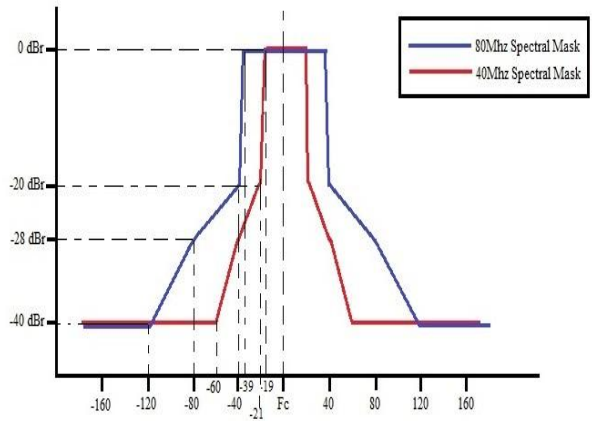

**Figure 4** IEEE 802.11ac Spectral Mask for 40 MHz & 80MHz measured with 100 kHz resolution bandwidth and 30 kHz video bandwidth.

Here, dBr = dB relative to the maximum spectral density of the signal. It can be seen that the roll-off from the 0dBr to the -20dBr points still occurs over a 2 MHz bandwidth, the same bandwidth for the 40 MHz mask. This means that in terms of the percentage of the signal bandwidth, the roll-off is twice as steep over these points.

**Table 3** IEEE 802.11ac Phy Layer Specification

| Feature | Mandatory | Optional |
|---|---|---|
| Channel bandwidth | 20MHz, 40 MHz, 80 MHz | 160 MHz, 80+80 MHz |
| FFT size | 64, 128, 256 | 512 |
| Data subcarriers / Pilots | 52 / 4, 108 / 6, 234 / 8 | 468 / 16 |
| Modulation types | BPSK, QPSK, 16-QAM, 64-QAM | 256-QAM |
| Spatial streams & MIMO | 1 | 2 to 8, TX beamforming, STBC, Multi-user-MIMO |

### 4.4. IEEE 802.11af

The IEEE 802.11af, also known as White-fi is meant to operate in the TV White Spaces, that will be the spectrum already allocated to the TV broadcasters however, not used at a certain location and time frame [12]. It uses cognitive radio technology to spot white spaces it could use. However, this cognitive technology will soon be predicated on an official geolocation database. This database can provide information on which frequency, at what time and under what conditions networks may operate. Table 4 summarizes the key features of this amendment.

**Table 4** Features of IEEE 802.11af

| Parameter | Details |
|---|---|
| Operating frequency range | 470-510 MHz |
| Channel bandwidth | 6 MHz |
| Transmission Power | 20 dBm |
| Modulation format | BPSK |
| Antenna Gain | 0dBi |

### 4.5. IEEE 802.11ah

The IEEE 802.11ah is directed at developing an international WLAN network which will allow user to gain access to sub carrier frequencies below 1GHz in the ISM band. One of the goals of this standard is to ensure that the transmission, ranges up to 1 km. It will even enable devices on the basis of the IEEE 802.11 standards to access short burst data transmissions like meter data. Additionally it can provide improve coverage range that will allow new applications such as, for example wide area based sensor networks, sensor backhaul systems and potential Wi-Fi off-loading functions to emerge [48]. This standard is under development and is predicted to be finalized by 2016.

### 4.6. IEEE 802.11ai

The IEEE 802.11ai is a forthcoming standard predicted to be finalized by 2015. It'll add a fast initial link setup (FILS) that might enable an STA to reach a protected link setup that will be significantly less than 100 ms [49]. An effective link setup process will then permit the STA to send IP traffic with a valid IP address through the AP.

### 4.7. IEEE 802.11mc

The IEEE 802.11mc resembles the IEEE 802.11m and is also scheduled to appoint an operating group with the job of maintenance of the standard around 2015.

### 4.8. IEEE 802.11aj

The IEEE 802.11aj is intended provide modifications to the IEEE 802.11ad Physical (PHY) layer and the Medium Access Control (MAC) layer for operation in the Chinese Milli-Meter Wave (CMMW) frequency bands like the 59-64 GHz frequency band. The amendment can also be meant to maintain backward compatibility with 802.11ad when it operates in the 59-64 GHz frequency band. The amendment shall also define modifications to the PHY and MAC layers allowing the operation in the Chinese 45 GHz frequency band. This standard is scheduled to be finalized by the end of 2016.

### 4.9. IEEE 802.11aq

The WLAN is fast evolving and is no more one, where stations are merely looking for just usage of internet service. This creates opportunities to supply new services, because the IEEE 802.11 standard must be enhanced to advertise and describe these new services.

The IEEE 802.11aq can provide mechanisms that will assist in pre-association discovery of services by addressing the methods to advertise their existence and enable delivery of information that describes them. These records about services will be made available ahead of association by stations operating on IEEE 802.11 wireless networks. This standard is scheduled to be published by 2015.

## 5. Issues with IEEE 802.11

Several extensions of the initial IEEE 802.11 standard have been developed but nonetheless there are lots of problems connected with the standards which are still needed to be addressed.

### 5.1. Security

Among the prime concerns in wireless networking is security. As WLANs operate within the shared medium, eavesdropping by unauthorized people and important information might be accessed with the utilization of malicious technologies. The original standard WEP had security flaws which lead the Wi-Fi forum to implement another encryption system WPA and later WPA2. Although WPA and WPA2 is a lot safer and provides good protection still it's not secure enough to be content with. More complicated encryption algorithms have to be implemented without decreasing the MAC layer throughput.

### 5.2. Data Rate

Another serious drawback of IEEE 802.11 wireless network is the data rate, which is quite low when compared with its wireline IEEE 802.3 counterpart. IEEE 802.11g supplies a data rate of 54Mbps. There is significant improvement in the information rates supplied by the IEEE 802.11n extension. The WWise and TGnSync proposal supported data rates upto 540Mbps and 630Mbps respectively. However, the 40 MHz channel required to guide such data rates aren't for sale in many countries. The more recent IEEE 802.11ac aims at providing data rates upto 7 Gbps. But this data rate is even still quite low as set alongside the 10 GHz data rate of the wireline network standard 802.3n.

## 6. Conclusion

In this paper we provided a detailed discussion on all of the reported upcoming IEEE 802.11 standards. Although some of the key issues have already been revised in the upcoming standards, some of the open issues still need to be addressed. The Task groups for the standards have already been set up and some of them have also succeeded in implementing the tasks it was assigned. Still, the IEEE 802.11 standard needs much development to give strong competition to its wireline counterpart i.e. IEEE 802.3 Ethernet.